\documentclass[aps,prb,twocolumn,superscriptaddress]{revtex4} 
\usepackage[dvipdfmx]{graphicx}
\usepackage[]{graphicx,color}
\usepackage[version=3]{mhchem}
\usepackage{bm} 
\usepackage{physics} 
\usepackage{ulem}

\begin{document}

\title{Superconducting LaPtH$_6$ with triatomic hydrogen units}



\author{Takahiro Ishikawa}%
 \email{takahiro.ishikawa@phys.s.u-tokyo.ac.jp}
 \affiliation{%
 Department of Physics, The University of Tokyo, 7-3-1 Hongo, Bunkyo-ku, Tokyo 113-0033, Japan
 }%
\author{Yuta Tanaka}%
 \affiliation{%
 Central Technical Research Laboratory, ENEOS Corporation, 8, Chidoricho, Naka-ku, Yokohama, Kanagawa 231-0815, Japan
  }%
\author{Shinji Tsuneyuki}%
 \affiliation{%
 Department of Physics, The University of Tokyo, 7-3-1 Hongo, Bunkyo-ku, Tokyo 113-0033, Japan
  }%
   
\date{\today}

\begin{abstract}
To veryfy ``hot superconductivity'' recently proposed in lanthanum hydride-based compounds, we explored thermodynamically stable and superconducting phases in the lanthanum (La)-platinum (Pt)-hydrogen (H) ternary system at 20\,GPa using an evolutionary construction scheme of a formation-enthalpy convex hull, universal neural network potential calculations, and density functional theory calculations. Although we found no evidence of the hot superconductivity in this ternary system, we predicted a unique compound, LaPtH$_6$, which has equilateral triangular H$_3$ units nearly forming a two-dimensional kagome lattice between La and Pt layers and shows the superconductivity at 18.67\,K. This structure is dynamically stable from ambient pressure to at least 200\,GPa and the superconducting critical temperature increases from 13.51 to 40.63\,K.  
\end{abstract}


\maketitle


\section{Introduction}

Metallic hydrides have attracted much attention as potential candidates for room-temperature superconductivity~\cite{Ashcroft1968,Ashcroft2004}. 
In 2015, high-temperature superconductivity was observed in sulfur hydride (S-H) under high pressure, and the superconducting critical temperature ($T_{\rm c}$) reaches 203\,K at pressure of 155\,GPa. This discovery triggered further theoretical studies on stoichiometry, crystal structure, and superconductivity in the S-H system under high pressure~\cite{Duan2015,Errea2015,Akashi2015,Errea2016,Ishikawa2016-SciRep,Akashi2016}, and it was clarified that H$_3$S compound emerges in the high $T_{\rm c}$ phase~\cite{Einaga2016}. Moreover, in 2018 and 2019, higher $T_{\rm c}$ was experimentally discovered in lanthanum hydride (La-H) under high pressure. First, LaH$_{10}$ compound was predicted to emerge as a stable phase showing high $T_{\rm c}$~\cite{Peng2017,Liu2017-LaH10}, and then it was experimentally confirmed to show $T_{\rm c}$ of 250--260\,K at 170--190\,GPa~\cite{Somayazulu2019,Drozdov2019-LaHx}. 
In addition, it is also interesting how the H atoms behave in crystal structure. In H$_3$S, the H$_2$ molecules are completely dissociated and form covalent bonds with the S atoms. On the other hand, in LaH$_{10}$, the H$_2$ molecules are weakly connected and form a cage structure surrounding the La atom. 

Many first-principles calculations have predicted the superconductivity in binary hydrides under high pressure with respect to more than 60 elements~\cite{Ishikawa2019}, whereas room-temperature superconductivity has not been obtained in the binary hydrides. 
Threfore, ternary hydrides are gathering attention as the next stage for the exploration of higher $T_{\rm c}$. In 2019, Sun \textit{et al.} predicted from first-principles calculations that \ce{Li_{2}MgH_{16}} shows $T_{\rm c}$ of 475\,K at 250\,GPa~\cite{Sun2019}. In 2020, Sun \textit{et al.} and Cui \textit{et al.} predicted that the insertion of CH$_{4}$ into H$_{3}$S causes the dynamical stabilization of the high $T_{\rm c}$ phase at lower pressures and \ce{CSH_{7}} shows the $T_{\rm c}$ values of 100--190\,K in pressure range of 100--200\,GPa~\cite{Sun2020_CHS,Cui2020_CSH}. 

Inspired by these theoretical results, experimentalists have explored novel ternary or multinary hydrides showing higher $T_{\rm c}$ at lower pressures. 
In 2022, Grockowiak \textit{et al.} reported ``hot superconductivity'' in ternary or multinary compounds based on the La-H system~\cite{Grockowiak2022}. The authors claim that $T_{\rm c}$ is increased to 550\,K by subsequent thermal excursion to high temperatures, which might be induced by the reaction of La-H with other materials existing in the sample chamber of the diamond anvil cell (DAC). 
Their experimental condition indicates that the candidates for the other materials are nitrogen (N) and boron (B) derived from ammonia borane used as hydrogen source, platinum (Pt), gold (Au), and gallium (Ga) from electrodes, etc. However, in first-principles calculations, the evidence of the hot superconductivity has not been found in the La-N-H, La-B-H, and La-Ga-H systems in high pressure region above 100\,GPa~\cite{Ge2021,DiCataldo2022} and the La-N-H system at lower pressure of 20\,GPa~\cite{Ishikawa2024_La-N-H}. 

In this paper, assuming that La-H is reacted with Pt, we searched for thermodynamically stable and metastable phases and superconducting phases in the La-Pt-H ternary system using an evolutionary construction scheme (ECS) of a formation-enthalpy convex hull, universal neural network potential calculations, and density functional theory calculations. 
We set the pressure value at 20\,GPa, which is easily generated by multi-anvil apparatus widely used in the field of materials science. Since the multi-anvil apparatus can accommodate much larger samples than DAC, synthesis of hydrides, x-ray diffraction measurements, superconductivity measurements (zero resistance and Meissner effect), and data analysis are much easier than those in DAC. 
Therefore, calculated results can immediately be validated by many experimental groups. 

\section{Computational details}

The experimental findings of the metallic hydrides with high $T_{\rm c}$ have all been motivated by theoretical crystal structure predictions~\cite{Oganov06,Oganov2010-variablecomposition,Wang2010,Pickard2011}, whereas the search for thermodynamically stable phases in ternary systems is computationally expensive. 
To overcome this problem, we performed fast and accurate search for stable phases by combining ECS of a formation-enthalpy convex hull~\cite{Ishikawa2020-CH}, universal neural network potential calculations by Matlantis~\cite{Matlantis}, and density functional theory calculations by the Quantum ESPRESSO (QE) code~\cite{QE}. The details of the method are given in Ref. [\onlinecite{Ishikawa2024_La-N-H}]. 

For the calculations by QE, we used the generalized gradient approximation by 
Perdew, Burke and Ernzerhof~\cite{PBE} for the exchange-correlation functional, and 
the Rabe-Rappe-Kaxiras-Joannopoulos ultrasoft pseudopotential~\cite{RRKJ90}. 
The energy cutoff was set at 80\,Ry for the wave function and 640\,Ry for the charge density. 
We increased the number of the $k$-point samplings in the Brillouin zone for optimization until the formation enthalpy is sufficiently converged. 
For the calculations on Matlantis, we used the PreFerred Potential (PFP) v.3.0.0~\cite{Takamoto2022}. and the L-BFGS optimization algorithm~\cite{LBFGS}. 
We performed the constant-pressure variable-cell optimization at 20\,GPa for the created structures. 
The space groups of the predicted structures were assigned using FYNDSYM~\cite{FINDSYM}. 

The superconducting $T_{\text{c}}$ was calculated using the Allen-Dynes modified McMillan formula~\cite{Allen-Dynes}, 
\begin{equation}
\label{AllenDynes}
T_{\text{c}}=\frac{f_{1}f_{2}\omega_{\log}}{1.2}
 \exp \left[ - \frac{ 1.04(1+\lambda )}
{ \lambda-\mu^{\ast}(1+0.62\lambda) } \right].
\end{equation}
The parameters, $\lambda$ and $\omega_{\log}$, are electron-phonon coupling constant and 
logarithmic-averaged phonon frequency, respectively, 
which represent a set of characters for the phonon-mediated superconductivity. 
$f_{1}$ and $f_{2}$ are correlation factors for the systems showing large $\lambda$. 
To obtain these parameters, we performed the phonon calculations implemented in the QE code. The effective screened Coulomb repulsion constant $\mu^{*}$ 
was assumed to be 0.10, which has been considered to be a reasonable value 
for hydrides. 
The $k$- and $q$-point grids for the calculations are listed in Table S1 in the Supplemental Material (SM)~\cite{SM_LaPtH}.

\section{Results}

\begin{figure}
\includegraphics[width=8.7cm]{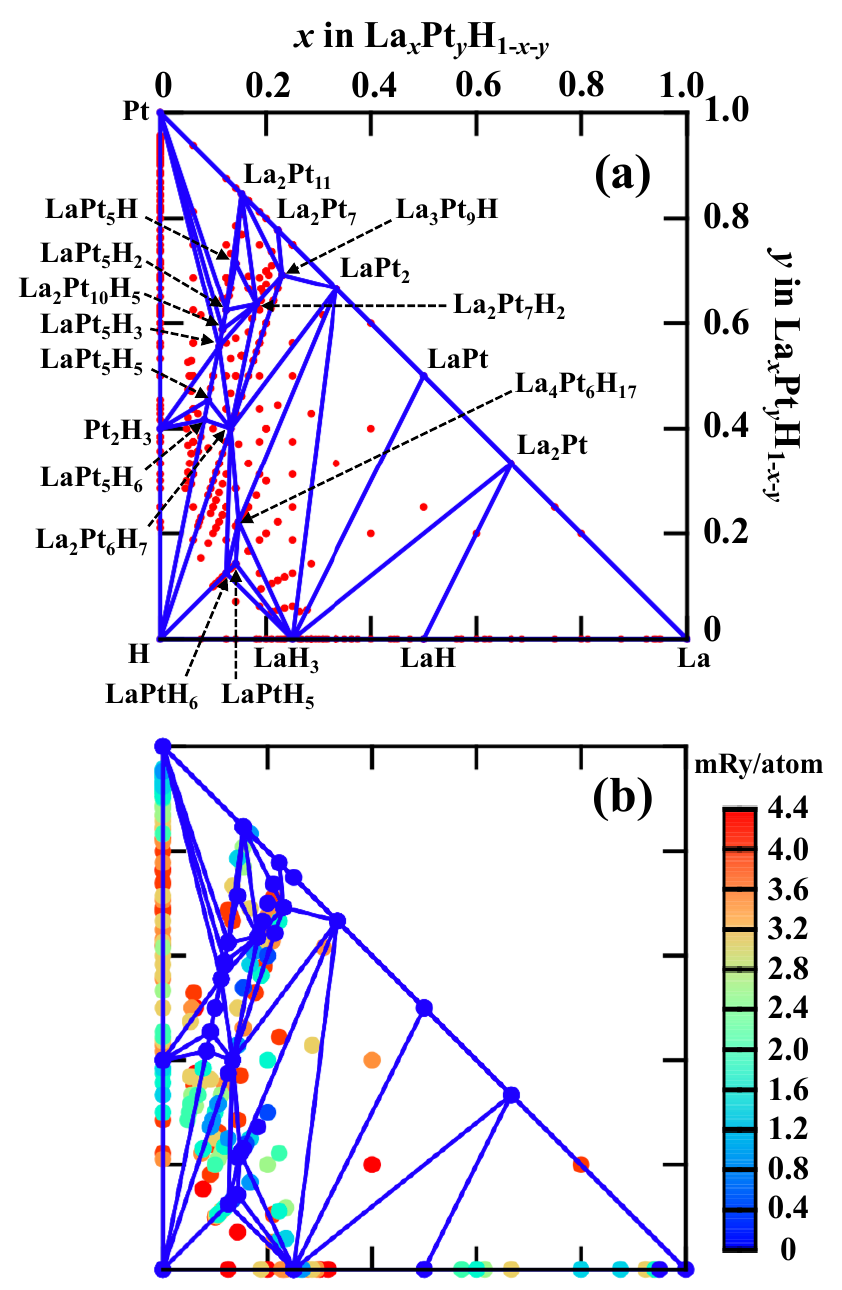}
\caption{\label{Fig_convexhull_QE} 
Formation-enthalpy convex hull diagram of the La-Pt-H system (\ce{La_{$x$}Pt_{$y$}H_{$1-x-y$}}) at 20\,GPa, obtained by QE. The convex hull is projected on the $xy$ plane, and the vertices (the intersections of the lines) correspond to thermodynamically stable compounds: (a) All the compounds investigated in this study and (b) moderately metastable compounds with $\Delta H$ less than 4.4\,mRy/atom.}
\end{figure}
Figure \ref{Fig_convexhull_QE}a shows the convex hull diagram of the formation enthalpy for the La-Pt-H system (\ce{La_{$x$}Pt_{$y$}H_{$1-x-y$}}, $0 \leq x \leq 1$, $0 \leq y \leq 1$) at the 10th generation obtained by the QE optimization after the screening by Matlantis (See Figs. S1 and S2 in SM for the convex hull diagram obtained by Matlantis only~\cite{SM_LaPtH}). In this figure, the convex hull is projected on the $xy$ plane, viewed along the $z$ axis showing the formation enthalpy. The lines and their intersections show the edges and vertices of the convex hull, respectively: the vertices correspond to thermodynamically stable compounds at 20\,GPa. The small dots are compositions created by ECS. We obtained twenty thermodynamically stable compounds as shown in the figures. 
Figure \ref{Fig_convexhull_QE}b shows only the moderately metastable compounds whose enthalpy difference ($\Delta H$) to the convex hull is less than 4.4\,mRy/atom. According to a large-scale data-mining study of the Materials Project in Ref. [\onlinecite{Sun2016}], metastable materials with $\Delta H < 70$\,meV/atom = 5.15\,mRy/atom predicted by DFT calculations are promising candidates to be synthesized by experimental techniques. 
Hence, we set the upper limit of $\Delta H$ at 4.4\,mRy/atom. The compounds with small $\Delta H$ are concentrated in the Pt-rich region including the line connecting between LaPt$_5$H$_6$ and La$_2$Pt$_{11}$. 

Next we compare the results obtained by ECS with the experimental and theoretical ones reported earlier. 
On the binary La-H line, we have already confirmed that the experimental results are well reproduced by our search~\cite{Ishikawa2024_La-N-H}. On the binary Pt-H line, 
previous first-principles calculations predict that PtH compound with a tetragonal $I\bar{4}m2$ structure is stabilized against decomposition into Pt and H at 21\,GPa~\cite{PhysRevLett.107.117002}. On the other hand, our calculations predict that PtH takes an orthorhombic $Imm2$ structure as the most stable one, whose enthalpy is lower by 1.1\,mRy/atom than that of $I\bar{4}m2$ at 20\,GPa. 
However, PtH is thermodynamically unstable at 20\,GPa because Pt$_2$H$_3$ with a trigonal $R\bar{3}m$ structure emerges as a vertex of the convex hull. Experimental studies show that unknown Pt-H phases emerge at 21\,GPa and a single phase of PtH with a hexagonal $P6_{3}/mmc$ structure is obtained at 42\,GPa~\cite{PhysRevB.83.214106}. Therefore, if Pt$_2$H$_3$ is considered as a phase of the unknown phases observed in the range of 21--42\,GPa, our results seem to be consistent with the experimental ones. 

Following the convex hull diagram predicted by ECS, we investigated the superconductivity for some compounds which are metallic and dynamically stable. While the La-Pt-H system basically shows weak superconductivity at 20\,GPa (see Table S2 and Fig. S3 in SM~\cite{SM_LaPtH}), four ternary compounds in hydrogen rich region show the superconducting $T_{\rm c}$ values above 10\,K. 
\begin{figure}
\includegraphics[width=8.7cm]{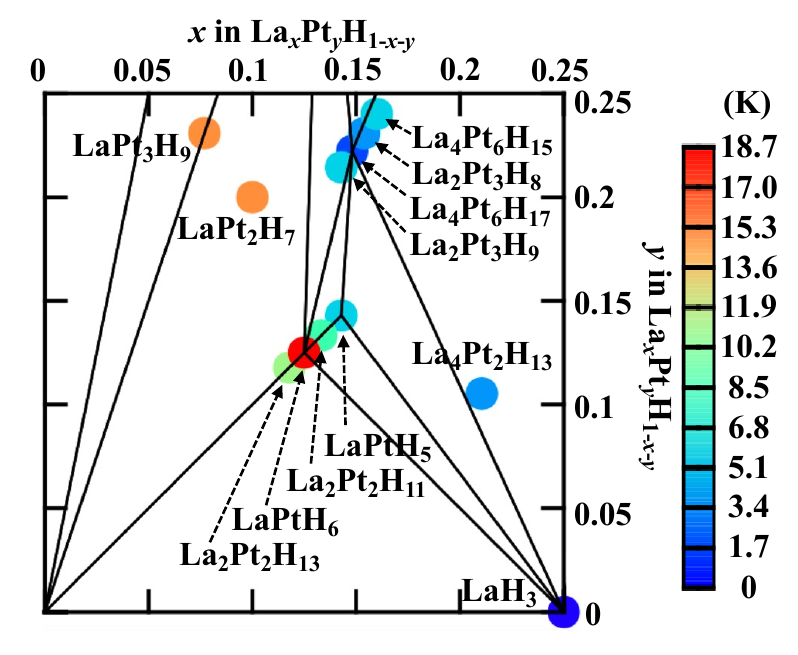}
\caption{\label{Fig_Tc_closeup} 
Superconducting $T_{\rm c}$ data for stable and moderately metastable compounds in hydrogen rich region at 20\,GPa. The color of the circle represents the magnitude of the $T_{\rm c}$ value.}
\end{figure}
\begin{table}
\caption{\label{Tcdata}
Superconductivity data of ternary compounds in the La-Pt-H system, predicted by the evolutionary construction scheme of a formation-enthalpy convex hull. $\Delta H$, $\lambda$, $\omega_{\log}$, and $T_{\rm c}$ are the enthalpy difference to the convex hull in the unit of mRy/atom, the electron-phonon coupling constant, 
the logarithmic-averaged phonon frequency in the unit of K, and the superconducting critical temperature in the unit of K, respectively.}
\begin{ruledtabular}
\begin{tabular}{ccccccc}
   La:Pt      & Compoud & Structure &  $\Delta H$ &  $\lambda$   &  $\omega_{\log}$  & $T_{\rm c}$ \\
\hline
  1:1            &  La$_2$Pt$_2$H$_{13}$ & $P1$ & 0.74 & 0.6150 & 449.5 & 11.46 \\
               &  LaPtH$_6$ & $R\bar{3}m$ & 0 & 0.7014 & 511.7 & 18.67 \\
              &  La$_2$Pt$_2$H$_{11}$ & $Cm$ & 0.78 & 0.5823 & 463.3 & 9.98 \\
              &  LaPtH$_5$ & $C2/m$ & 0 & 0.4942 & 551.6 & 6.56 \\
  2:3            &  La$_2$Pt$_3$H$_9$ & $P1$ & 0.34 & 0.5569 & 343.3 & 6.38 \\
              &  La$_4$Pt$_6$H$_{17}$ & $P1$ & 0 & 0.4364 & 379.3 & 2.58 \\
              &  La$_2$Pt$_3$H$_8$ & $P1$ & 0.12 & 0.5217 & 307.3 & 4.52 \\
              &  La$_4$Pt$_6$H$_{15}$ & $P1$ & 0.70 & 0.6039 & 265.5 & 6.40 \\
    1:2          &  LaPt$_2$H$_7$ & $P1$ & 2.11 & 0.7879 & 305.3 & 14.59 \\
   1:3           &  LaPt$_3$H$_9$ & $P1$ & 1.95 & 0.8158 & 284.0 & 14.61 \\
  2:1            & La$_{4}$Pt$_{2}$H$_{13}$ & $P1$ & 3.78 & 0.5277  & 324.9  & 4.98 \\
\end{tabular}
\end{ruledtabular}
\end{table}
Figure \ref{Fig_Tc_closeup} shows the $T_{\rm c}$ values in hydrogen rich region, $0 \leq x \leq 0.25$ and $0 \leq y \leq 0.25$. The superconducting compounds are classified into five types following the ratio of La and Pt: (i) 1:1, (ii) 2:3, (iii) 1:2, (iv) 1:3, and (v) 2:1 types. LaPtH$_6$ of the 1:1 type shows the highest $T_{\rm c}$ of all the compounds predicted in this study. The $T_{\rm c}$ value is 18.67\,K ($\lambda = 0.7014$ and $\omega_{\log} = 511.7$\,K) and is higher by about 4\,K than the highest $T_{\rm c}$ in the La-N-H system at the same pressure predicted earlier~\cite{Ishikawa2024_La-N-H}. As listed in Table \ref{Tcdata}, the superconductivity is weakened by increase or decrease of the hydrogen concentration of LaPtH$_6$ within the 1:1 type. LaPt$_2$H$_7$ of the 1:2 type and LaPt$_3$H$_9$ of the 1:3 type show the $T_{\rm c}$ value of about 14.6\,K, which is the second highest $T_{\rm c}$ of all the predicted compounds. The 2:3 and 2:1 types show the $T_{\rm c}$ values from 2.58 through 6.40\,K. The structure data (CIF files) of all the superconducting compounds are provided in SM~\cite{SM_LaPtH}. 

\begin{figure*}
\includegraphics[width=15cm]{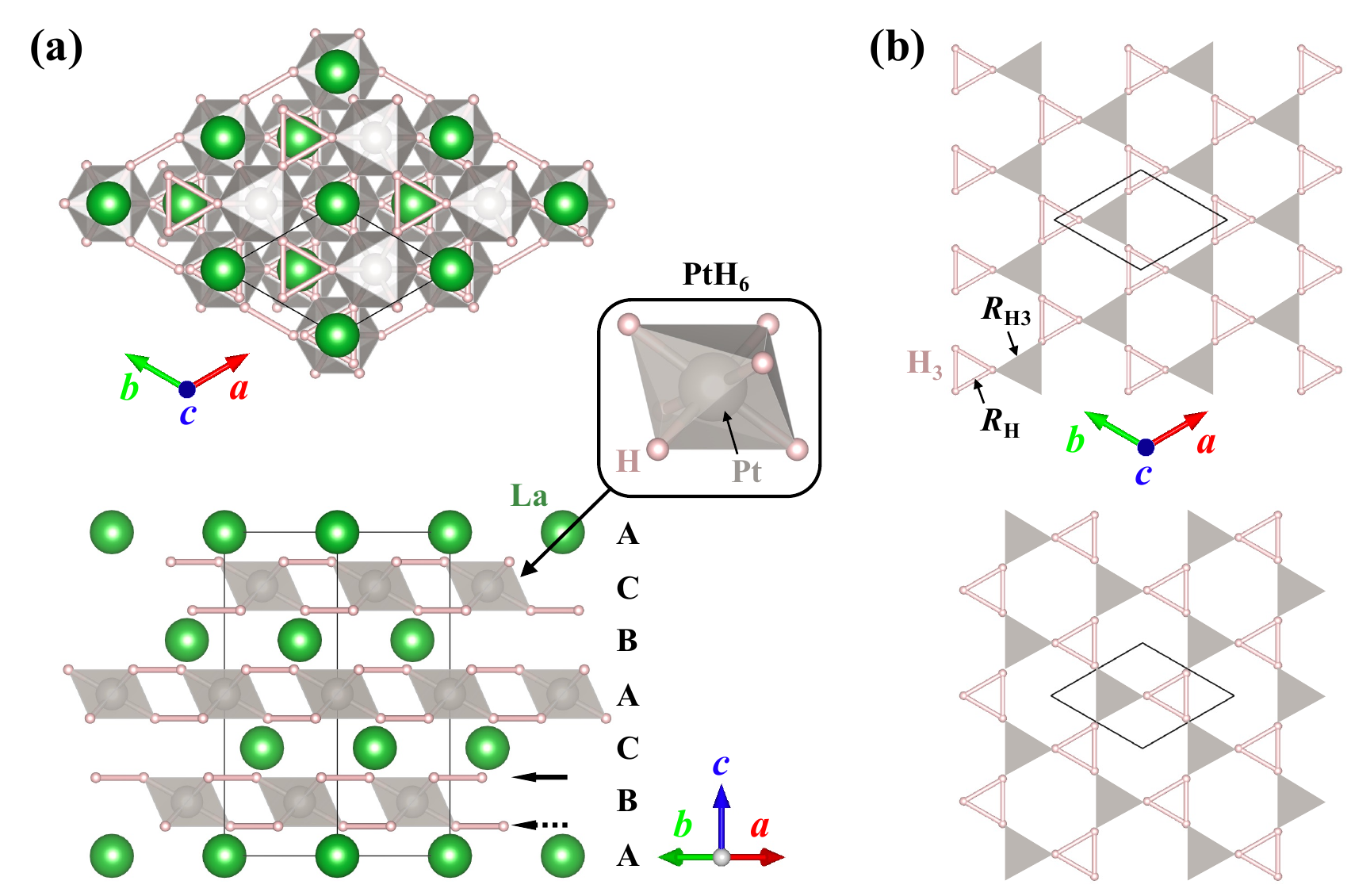}
\caption{\label{Fig_LaPtH6} 
(a) Crystal structure of LaPtH$_6$, assigned as a trigonal $R\bar{3}m$. Octahedral PtH$_6$ units are formed, and the La layer and the PtH$_6$ layer are stacked along the $c$ axis with the order of A(La)-B(PtH$_6$)-C(La)-A(PtH$_6$)-B(La)-C(PtH$_6$). The solid lines show the unit cell.  (b) Views of the H layers. The upper and lower figures correspond to the layers indicated by the solid and dotted horizontal arrows in (a), respectively. The H$_3$ units nearly form two-dimensional kagome lattices, in which $R_{\text{H}}$ and $R_{\text{H3}}$ represent the H-H bond length of the H$_3$ unit and the nearest neighbor distance between the H$_3$ units, respectively. The shaded triangles show the facets of the octahedron. The structure was drawn with VESTA~\cite{VESTA}.}
\end{figure*}

Here, we focus on LaPtH$_6$ showing the highest $T_{\rm c}$. This compound has the highest concentration of hydrogen of all the stable compounds predicted in this study and forms a simple but unique crystal structure. Figure \ref{Fig_LaPtH6}a shows the structure of LaPtH$_6$, assigned as a trigonal $R\bar{3}m$. 
This structure has octahedral PtH$_6$ units, and Bader charge analysis shows that the La atom and the PtH$_6$ unit are positively and negatively charged, respectively (Table S3 and Fig. S4 in SM). The positively charged La ions and the negatively charged PtH$_6$ units individually form triangular lattice layers, and they are stacked along the $c$ axis with the order of A(La)-B(PtH$_6$)-C(La)-A(PtH$_6$)-B(La)-C(PtH$_6$). Note that A, B, and C are the same as the layer positions describing the hexagonal close-packed structure (A-B) or the face-centered cubic structure (A-B-C) and a component forming each layer is shown in parentheses. As shown in Fig. \ref{Fig_LaPtH6}b, three H atoms form an equilateral triangle like so-called triatomic hydrogen, H$_3$. The H-H bond length of an H$_3$ unit ($R_{\text{H}}$) is 1.87{\AA} and the nearest neighbor distance between the H$_3$ units ($R_{\text{H3}}$) is 2.55{\AA}, in which the atomic alignment is similar to a two-dimensional kagome lattice. The center of the mass of an H$_3$ unit occupies the layer position other than those of the neighboring layers, e.g., if an H$_3$ unit is intercalated between the A(La) and B(Pt) layers, the center of the mass occupies the C position. In addition, the H$_3$ units in the same layer point the same direction: parallel or anti-parallel to the $\bm{a}-\bm{b}$ direction (Fig. \ref{Fig_LaPtH6}b). If the former and latter are defined as R and L, respectively, the structure of LaPtH$_6$ is also represented as the stacking structure with the order of A(La)-C(H$_3$:L)-B(Pt)-A(H$_3$:R)-C(La)-B(H$_3$:L)-A(Pt)-C(H$_3$:R)-B(La)-A(H$_3$:L)-C(Pt)-B(H$_3$:R) along the $c$ axis. In contrast, LaH$_3$ considered as an analog of LaPtH$_6$ has no H$_3$ unit, and the H atoms form an extensive covalent network with the bond length of 2.27{\AA} (Fig. S5 in SM~\cite{SM_LaPtH}). 

\begin{figure}
\includegraphics[width=8.7cm]{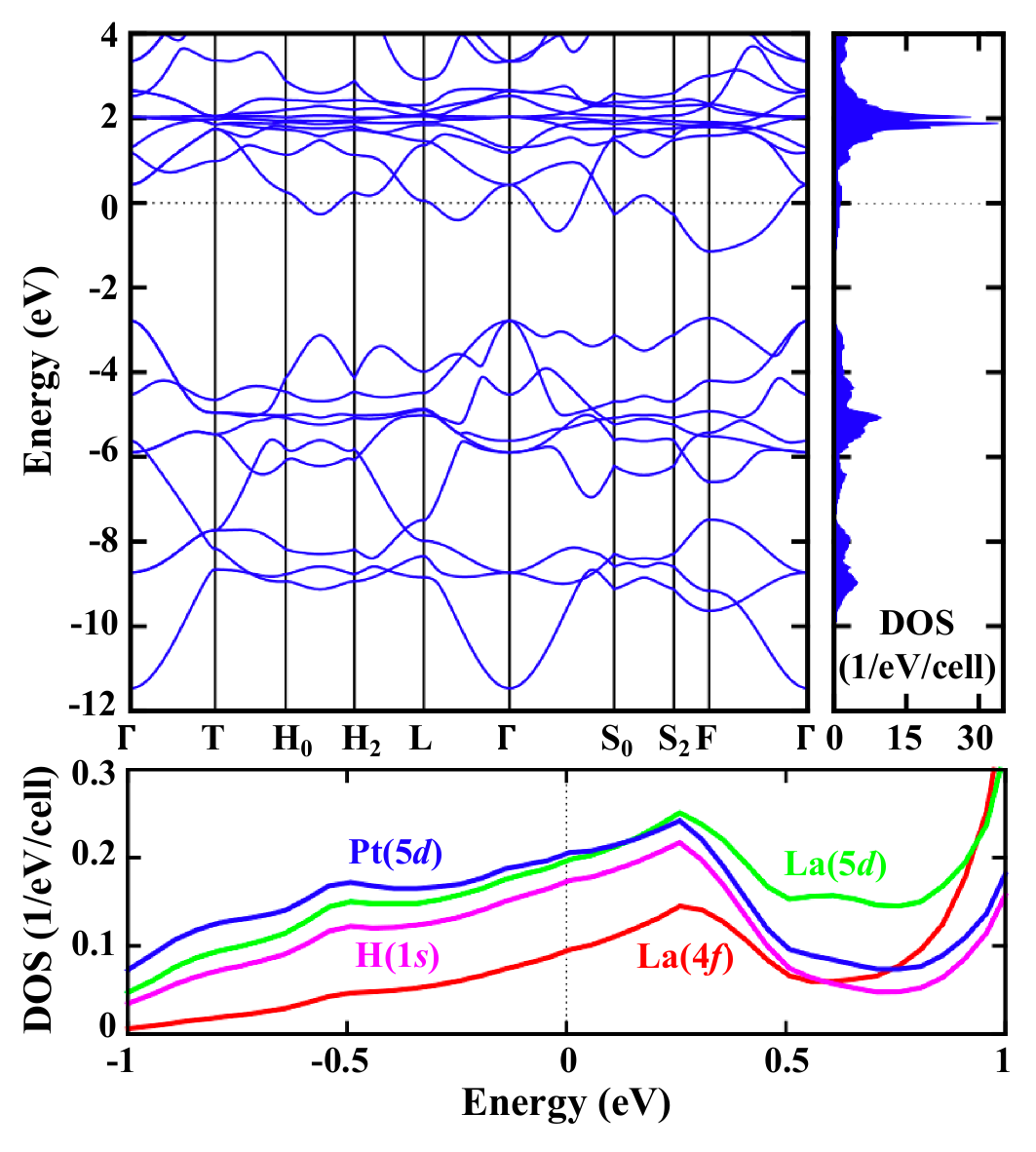}
\caption{\label{Fig_band} 
Electronic band dispersion and density of states (DOS) of LaPtH$_6$ at 20\,GPa (upper), and partial DOS in the vicinity of the Fermi level (lower). La($5d$), La($4f$), Pt($5d$), and H($1s$) represent the $5d$ states of La, the $4f$ states of La, the $5d$ states of Pt, and the $1s$ states of H, respectively. The Fermi level is set to zero. }
\end{figure}

Figure \ref{Fig_band} shows the electronic band structure and the density of states (DOS) for $R\bar{3}m$ LaPtH$_6$ at 20\,GPa. This compound has a band structure like electron-doped insulators, in which the Fermi level lies in the bands immediately above the gap. The bottom panel of this figure shows the partial DOS. The $5d$ states of La and Pt [La($5d$) and Pt($5d$)] dominantly contribute to the electronic states at the Fermi level. The hydrogen $1s$ [H($1s$)] also shows a large contribution to the electronic states at the Femi level, which causes the enhancement of the superconductivity. In hydrides based on lanthanides like LaH$_{10}$ at 250\,GPa, the $4f$ states of La [La($4f$)] dominantly contribute to the states at the Fermi level, which causes the high $T_{\rm c}$ superconductivity~\cite{Liu2017-LaH10,Plekhanov2022}. However, in the case of LaPtH$_6$,  the contribution of La($4f$) is smaller than those of La($5d$), Pt($5d$), and H($1s$). 
\begin{figure}
\includegraphics[width=8.7cm]{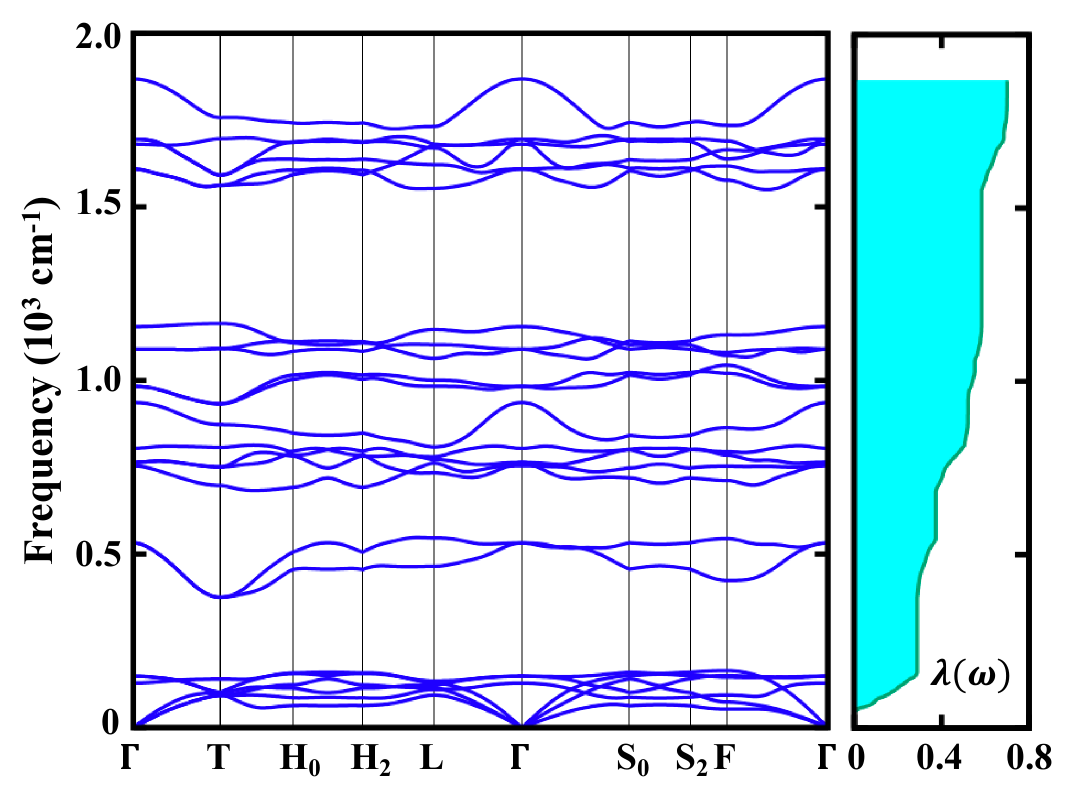}
\caption{\label{Fig_phonon} 
Phonon dispersion and cumulative electron-phonon coupling constant $\lambda(\omega)$ of LaPtH$_6$ at 20\,GPa. $\omega$ represents phonon frequency. }
\end{figure}
Figure \ref{Fig_phonon} shows the phonon dispersion and the cumulative electron-phonon coupling constant $\lambda(\omega)$ of LaPtH$_6$, where $\omega$ represents the phonon frequency. The low frequency modes below 164\,cm$^{-1}$ contribute to 41.2\% of the total $\lambda$ value, 0.7014. In addition, the high-frequency modes above 1550\,cm$^{-1}$, related to vibrations of the H$_3$ units, also contribute to 16.8\% of $\lambda$. The phonon of $R\bar{3}m$ LaPtH$_6$ is robust against pressure and has no imaginary frequency from 0 to at least 200\,GPa (Fig. S6 in SM~\cite{SM_LaPtH}). 
Thus, once $R\bar{3}m$ LaPtH$_6$ is synthesized under high pressure, it can be quenched to ambient pressure. 
$R_{\text{H}}$ is almost equal to $R_{\text{H3}}$ at 0\,GPa, in which the H atoms nearly form a complete two-dimensional kagome lattice. As pressure increases, the PtH$_6$ units approach each other and the size of the H$_3$ unit becomes smaller. In contrast, $R_{\text{H3}}$ increases to about 2.6{\AA} (Table \ref{Tc_pressure}). We calculated the $T_{\rm c}$ values and found that $T_{\rm c}$ increases from 13.51 to 40.63\,K as pressure increases from 0 to 200\,GPa (Table \ref{Tc_pressure}). These results suggest that LaPtH$_6$ can not explain the hot superconductivity experimentally observed at 180\,GPa. 
\begin{table}
\caption{\label{Tc_pressure}
Pressure ($P$) dependence of superconductivity for $R\bar{3}m$ LaPtH$_6$. The parameters $R_{\text{H}}$ and $R_{\text{H3}}$ show the H-H bond length of an H$_3$ unit and the nearest neighbor distance between the H$_3$ units in the same layer, respectively.}
\begin{ruledtabular}
\begin{tabular}{cccccc}
   $P$ (GPa)      & $R_{\text{H}}$ ({\AA}) & $R_{\text{H3}}$ ({\AA}) & $\lambda$   &  $\omega_{\log}$ (K)  & $T_{\rm c}$ (K) \\
\hline
0  & 2.23 & 2.43 & 0.7893 & 282.1 & 13.51 \\
20 & 1.87 & 2.55 & 0.7014 & 511.7 & 18.67 \\
 50  & 1.62 & 2.63 & 0.6628  & 771.6  & 24.32 \\
100  & 1.42 & 2.66 & 0.6856  & 942.4  & 32.43 \\
150  & 1.30 &2.66 & 0.7121  & 1016.7  & 38.48 \\
200  &1.23 & 2.64 & 0.7182  & 1051.3  & 40.63 \\
\end{tabular}
\end{ruledtabular}
\end{table}

\section{Discussions and conclusions}

To verify the hot superconductivity reported earlier, we explored thermodynamically stable and metastable phases in the ternary La-Pt-H system at 20\,GPa using ECS, the universal neural network potential calculations by Matlantis, and the DFT calculations by QE. While new superconducting compounds were predicted in the hydrogen-rich region, the highest $T_{\rm c}$ value is 18.67\,K obtained in LaPtH$_6$, which is higher by about 4\,K than the highest $T_{\rm c}$ in the La-N-H system at the same pressure. Although we found no evidence of the hot superconductivity in this study, we found LaPtH$_6$ with a simple but unique crystal structure, in which the equilateral triangular H$_3$ units almost form a two-dimensional kagome lattice between the La and Pt layers. This behavior of the H atoms is significantly different from the S-H and La-H systems cases: Dissociation of the H$_2$ molecules in S-H and weak connection of the H$_2$ molecules in La-H. We also found that the vibration modes of the H$_3$ units contribute to the enhancement of the superconductivity. Since the phonon of LaPtH$_6$ with the H$_3$ units is robust against pressure and shows no imaginary frequency over a wide range from ambient pressure to at least 200\,GPa, it is quite possible that the compound is synthesized by experiments and is quenched to ambient condition. 

The existence of triatomic hydrogen, H$_3$, was first proposed by Thomson in 1911~\cite{Thomson1911_H3}, and spectroscopic lines of neutral H$_{3}$ was first observed by Herzberg in 1979~\cite{Herzberg1979_H3} via many experimental and theoretical studies~\cite{Kragh2011_review}. 
A neutral H$_3$ molecule is unstable and breaks up in under a microsecond after the formation, 
and an H$_{3}^{+}$ ion is one of the most abundant molecular ions in the universe, which is believed to have played a crucial role in the cooling of early stars in the history of the universe. 
In addition, a kagome lattice consisting of the H atoms may exhibit novel physical properties connected with the geometrical frustration.
Thus, our predicted LaPtH$_6$ could be an excellent model for exploring these phenomena.

\begin{acknowledgments}
This work was supported by JSR-UTokyo Collaboration Hub, CURIE, and JSPS KAKENHI under Grant-in-Aid for 
Scientific Research (C) (23K03316), Scientific Research (B) (24K00544), and Scientific Research (S) (20H05644). A part of the computation was performed using the facilities of the Supercomputer Center, the Institute for Solid State Physics, the University of Tokyo.
\end{acknowledgments}

\begin{thebibliography}{43}
\bibitem[{SM_()}]{SM_LaPtH}
\bibinfo{note}{See Supplemtal Material at [URL will be inserted by publisher]
  for calculation accuracy of Matlantis and calculation conditions and
  superconductivity data for stable and moderately metastable compounds in the
  La-Pt-H system at 20{\,}GPa.}
\end{thebibliography}

\end{document}